\documentclass[english,aps,preprint]{revtex4}
\usepackage{times}
\usepackage[T1]{fontenc}
\usepackage[latin1]{inputenc}
\usepackage{amsmath}
\usepackage{color}
\usepackage{graphicx}
\usepackage{amssymb}

\makeatletter
 \newenvironment{lyxlist}[1]
   {\begin{list}{}
     {\settowidth{\labelwidth}{#1}
      \setlength{\leftmargin}{\labelwidth}
      \addtolength{\leftmargin}{\labelsep}
      }}
   {\end{list}}

\usepackage{babel}
\makeatother
\begin{document}

\preprint{This line only printed with preprint option}

\title{Dispersive destabilization of nonlinear light propagation in fiber
Bragg gratings}

\author{Carlos Martel}

\email{martel@fmetsia.upm.es}

\affiliation{Depto. de Fundamentos Matemáticos, E.T.S.I. Aeronáuticos, Universidad
Politécnica de Madrid, Plaza Cardenal Cisneros 3, 28040 Madrid, Spain }

\begin{abstract}
The effect of retaining the material dispersion terms in the nonlinear
coupled mode equations (NLCME) that describe light propagation in
fiber Bragg gratings is analyzed. It is found that dispersion is responsible
for new instabilities of the uniform states and gives rise to new
complex spatio temporal dynamics that is not captured by the standard
NLCME formulation. A detailed analysis of the effect of dispersion
on the linear stability characteristics of the uniform solutions is
presented and some numerical integrations of the NLCME with dispersion
are also performed in order to corroborate the theoretical results.
\end{abstract}
\maketitle

\section{introduction}

\textbf{The nonlinear coupled mode equations (NLCME) are the envelope
equations currently used to study the weakly nonlinear dynamics of
light propagation in fiber Bragg gratings. These are universal equations
in the sense that they describe the large scale pattern formation
of any conservative, spatially extended propagative system with a
weak periodic spatial structure. Material dispersion effects are systematically
neglected in the derivation that leads to the NLCME. This paper presents
a study of the effect of retaining the dispersive terms on the stability
of the uniform states and it concludes that these terms should be
considered because they give rise to new instabilities and dynamics
that are not captured by the NLCME. The NLCME fail, in general, to
predict the system evolution and the dispersive terms produce the
onset of short scale perturbations that evolve and spread over the
domain in the NLCME time scale. Numerical results of the integration
of the NLCME with dispersive terms are also provided to corroborate
the theoretical linear stability results. }

Fiber Bragg gratings (FBG), i.e., optical fibers with a periodic variation
of the refractive index along its length, exhibit frequency gaps where
light resonates with the periodic structure of the fiber and propagation
is not permitted. This frequency dependent reflection characteristic
together with the recent developments in grating fabrication techniques
has allowed for successful application of FBG in optical communications
and fiber sensing, see e.g. \cite{Kashyap99} for a recent review.
Apart from this purely linear effect, the combination of the nonlinearity
of the fiber that comes into play at sufficiently high intensities
with the spatially distributed reflection produced by the grating
gives rise to localized structures known as gap solitons (see \cite{deSterkeSipe94,Aceves00,KivsharAgrawal03}
and references therein). These gap solitons are solitary waves that,
at least in theory, can propagate through the FBG at any speed between
zero and the speed of light of the fiber without grating. This possibility
of capturing light (gap soliton propagation at zero speed) is currently
a topic of great interest because of its promising future applications
in the production of devices for optical storage of information.

Light propagation in a nonlinear FBG is usually analyzed using the
so-called nonlinear coupled mode equations (NLCME) \textcolor{black}{\cite{deSterkeSipe94,WinfulCooperman82},\begin{eqnarray}
 & A_{t}^{+}=\phantom{-}v_{g}A_{x}^{+}+\textrm{i}\Delta nA^{-}+\textrm{i}gA^{+}(\sigma|A^{+}|^{2}+|A^{-}|^{2}),\label{NLCME1}\\
 & A_{t}^{-}=-v_{g}A_{x}^{-}+\textrm{i}\Delta nA^{+}+\textrm{i}gA^{-}(\sigma|A^{-}|^{2}+|A^{+}|^{2}),\label{NLCME2}\end{eqnarray}
}which prescribe the time ($t$) evolution along the fiber ($x$)
of the complex envelopes $A^{+}$ and $A^{-}$ of the wavetrains that
resonate with the grating. All variables are assumed to have been
previously nondimensionalized in a similar way as in \cite{GoodmanWeinsteinHolmes01},
that is, space has been made nondimensional with the grating period
$\lambda_{\textrm{g}}$, time with $\lambda_{\textrm{g}}/c$ ($c$
is the vacuum speed of light), the grating amplitude with the mean
value of the refractive index, and the fields with the characteristic
size that results from the Kerr nonlinearity. The actual electric
field inside the FBG can be written, in first approximation, as the
superposition of two modulated, counterpropagating wavetrains\begin{equation}
E\sim\begin{array}{c}
A^{+}(x,t)\end{array}\textrm{e}^{\textrm{i}\pi x+\textrm{i}\omega_{0}t}+A^{-}(x,t)\textrm{e}^{-\textrm{i}\pi x+\textrm{i}\omega_{0}t}+\:\mbox{c.c.}+\cdots,\label{Efield}\end{equation}
 with wavenumber $\pi$ and frequency $\omega_{0}=\omega(k=\pi)$,
which is given by the dispersion relation of the fiber without grating
$\omega(k)$ (c.c. stands for the complex conjugate). The wavelength
of the wavetrains, $\lambda_{0}=2$, is twice the period of the grating,
that is assumed to have a spatial profile of the form $\cos(2\pi x)$.
This situation can be seen as a spatial parametric resonance that
couples both wavetrains: the interaction of one of the wavetrains
with the grating gives rise to a resonant term that excites the other
wavetrain,\[
\cos(2\pi x)A^{+}\textrm{e}^{\textrm{i}\pi x+\textrm{i}\omega_{0}t}\sim(\textrm{e}^{\textrm{i}2\pi x}+\textrm{e}^{-\textrm{i}2\pi x})A^{+}\textrm{e}^{\textrm{i}\pi x+\textrm{i}\omega_{0}t}\rightarrow A^{+}\textrm{e}^{-\textrm{i}\pi x+\textrm{i}\omega_{0}t}\quad\textrm{($A^{-}$ resonant), }\]
and viceversa.

The NLCME take into account the effect of (i) transport at the group
velocity, which is given by $v_{g}=d\omega/dk$ at $k=\pi$, (ii)
coupling of the wavetrains through the spatially extended reflection
produced by the grating, with $\Delta n$ proportional to the strength
of the grating, and (iii) cubic nonlinear interaction, where $\sigma=\frac{1}{2}$
as a result of the cubic Kerr nonlinearity of the original physical
problem for the electric field \cite{deSterkeSipe94}. The NLCME are
also used to describe the evolution of Bose-Einstein condensates in
optical lattices \cite{yulinskryabin03,SakaguchiMalomed04} and, in
general, the NLCME can be regarded as a normal form that applies to
any dissipationless propagative system, extended in one spatial dimension,
reflection and translation invariant, and with a spatial parametric
periodic variation.

The derivation of the NLCME is carried out using a standard multiple
scales procedure that requires to assume small amplitudes in (\ref{Efield})
with slow spatial and temporal dependence (slow as compared with the
wavelength and temporal period of the basic wavetrains), i.e., 

\begin{equation}
\cdots\ll|A_{xx}^{\pm}|\ll|A_{x}^{\pm}|\ll|A^{\pm}|\ll1,\quad\cdots\ll|A_{t}^{\pm}|\ll|A^{\pm}|\ll1,\label{Slowxt}\end{equation}
 \textcolor{black}{and small grating intensity, \begin{equation}
\Delta n\ll1.\label{Deltanll1}\end{equation}
 The rest of the parameters:} $\omega_{0}$, $v_{g}$ and $g$\textcolor{black}{,
are all considered order one quantities. A detailed derivation of
the NLCME from the 1D Maxwell-Lorentz equations can be found in \cite{GoodmanWeinsteinHolmes01}.
In the process of derivation of the NLCME material dispersion terms
($\sim\textrm{i}A_{xx}^{\pm}$) are systematically neglected against
transport terms ($\sim A_{x}^{\pm}$). Notice that this is not the
case in the derivation of the nonlinear Schrödinger equation (NLS)
for a single modulated wavetrain propagating in an uniform fiber,}
where both effects are considered because the transport effect can
be easily accounted for just by using a moving coordinate frame that
travels with the group velocity \cite{Hasegawa90}. If we retain the
material dispersion terms, then the resulting equations for $A^{\pm}$
\textcolor{black}{read\begin{eqnarray}
 & A_{t}^{+}=\phantom{-}v_{g}A_{x}^{+}+\textrm{i}dA_{xx}^{+}+\textrm{i}\Delta nA^{-}+\textrm{i}gA^{+}(\sigma|A^{+}|^{2}+|A^{-}|^{2}),\label{NLCMEd+}\\
 & A_{t}^{-}=-v_{g}A_{x}^{-}+\textrm{i}dA_{xx}^{-}+\textrm{i}\Delta nA^{+}+\textrm{i}gA^{-}(\sigma|A^{-}|^{2}+|A^{+}|^{2}),\label{NLCMEd-}\end{eqnarray}
where $d=-d^{2}\omega/dk^{2}$ at $k=\pi$ is the dispersion coefficient
that is, generically, an order one quantity. Note that the coefficients
of the linear terms of the above envelope equations can be easily
obtained from the coefficients of the power expansion of the dispersion
relation $\omega(k)$ at $k=\pm\pi$ (see e.g. the review \cite{crosshohenberg93}).}

\textcolor{black}{In the above equations the transport terms cannot
be removed using a moving coordinate transformation and the two effects,
transport and dispersion, with different asymptotic orders, see eq.
(\ref{Slowxt}), must be considered simultaneously and give rise to
two different slow spatial scales:}

\begin{lyxlist}{0}
\item [a)]A transport scale, $\delta x_{\textrm{trans}}\sim1/\Delta n\gg1$,
which comes from the balance between the transport along the fiber
at the group velocity and the continuous spatial reflection produced
by the grating and is the slow spatial scale used in the NLCME derivation
\cite{WinfulCooperman82,deSterkeSipe94,GoodmanWeinsteinHolmes01}.
\item [b)]And a dispersive scale, $\delta x_{\textrm{disp}}\sim1/\sqrt{\Delta n}\gg1$,
which results from the balance between dispersion and grating reflection.
This dispersive scale is small as compared with the transport scale
but still large as compared with the basic wavetrain wavelength, i.e.,\[
\delta x_{\textrm{trans}}\gg\delta x_{\textrm{disp}}\gg1,\]
 and corresponds to the standard dispersive scaling used in the derivation
of the NLS equation \cite{Hasegawa90}.
\end{lyxlist}
\textcolor{black}{If we look for states exhibiting only transport
scales then the dispersion terms in equations (\ref{NLCMEd+})-(\ref{NLCMEd-})
can be neglected in first approximation and these transport dominated
solutions are thus well described by the NLCME. But, in general, the
question of whether the system will develop dispersive scales or not
is a stability question and its answer is not known a priori. }

The purpose of the present paper is to demonstrate that the material
dispersion terms in eqs. \textcolor{black}{(\ref{NLCMEd+})-(\ref{NLCMEd-})}
should not, in general, be neglected because the dispersive scales
can be unstable and grow and spread all over the domain. In other
words, we will show that stable solutions of the NLCME can be unstable
in the context of the more general problem given by eqs. \textcolor{black}{(\ref{NLCMEd+})-(\ref{NLCMEd-}),}
and can develop dispersive scales that give rise to very complicated
spatio-temporal states that are not included in the NLCME description. 

To this end, we will consider equations \textcolor{black}{(\ref{NLCMEd+})-(\ref{NLCMEd-})}
with simple periodic boundary conditions, \[
A^{\pm}(x+L,t)=A^{\pm}(x,t),\qquad\textrm{with }L\gg1,\]
which correspond to a ring shaped FBG with length $L$ containing
an even number of grating periods (like the fiber-loops considered
in the simulations of multiple gap soliton collisions in \cite{MakMalomedChu03}),
and we will focus on the effect of dispersion on the stability of
the simplest solutions of the NLCME, namely, the solutions with constant
uniform modulus. Equations \textcolor{black}{(\ref{NLCMEd+})-(\ref{NLCMEd-}})
can be further simplified if we rescale them with the transport scales
commonly used for the NLCME \cite{WinfulCooperman82,deSterkeSipe94,GoodmanWeinsteinHolmes01},\begin{equation}
\tilde{t}=tv_{g}/L,\quad\tilde{x}=x/L,\quad\tilde{{A^{\pm}}}=\sqrt{gL/v_{g}}A^{\pm},\quad\textrm{and}\quad\kappa=\Delta nL/v_{g},\label{escalados}\end{equation}
 where the scaled grating strength $\kappa\sim1$, i.e., the FBG length
is of the order of the transport scale $\delta x_{\textrm{trans}}$.
Note that $g$ can always be made positive by making, if necessary,
the change of variables $A^{\pm}\rightarrow\pm A^{\pm}$ and taking
complex conjugates in eqs. \textcolor{black}{(\ref{NLCMEd+})-(\ref{NLCMEd-}).}
The resulting nonlinear coupled mode equations with dispersion (NLCMEd),
after dropping tildes, \textcolor{black}{take the form\begin{eqnarray}
 & A_{t}^{+}=\phantom{-}A_{x}^{+}+\textrm{i}\varepsilon A_{xx}^{+}+\textrm{i}\kappa A^{-}+\textrm{i}A^{+}(\sigma|A^{+}|^{2}+|A^{-}|^{2}),\label{disp+}\\
 & A_{t}^{-}=-A_{x}^{-}+\textrm{i}\varepsilon A_{xx}^{-}+\textrm{i}\kappa A^{+}+\textrm{i}A^{-}(\sigma|A^{-}|^{2}+|A^{+}|^{2}),\label{disp-}\\
 & A^{\pm}(x+1,t)=A^{\pm}(x,t).\label{dispcc}\end{eqnarray}
where the scaled dispersion $\varepsilon=d/(Lv_{g})\ll1$ and $\kappa\sim1$,
as mentioned above. With this new scaling the transport and dispersive
scales are now $\delta x_{\textrm{trans}}\sim1$ and $\delta x_{\textrm{disp}}\sim\sqrt{|\varepsilon|}\ll1$,
and the wavelength of the basic wavetrains in (\ref{Efield}) is $\sim|\varepsilon|\ll1$.
The small dispersion term $\varepsilon\ll1$ in the above equations
is essential because it takes into account the fact that we are simultaneously
considering two effects, transport and dispersion, that have different
asymptotic orders due to the original slow modulation assumption (\ref{Slowxt}).
There are some previous works \cite{ChampneysMalomedFriedman98,ChampneysMalomed99,SchollmannMayer00}
that study the structure and stability of the localized solutions
of the NLCMEd (\ref{disp+})-(\ref{disp-}), but they all consider
$\varepsilon\sim1$ and do not study the small dispersion limit $\varepsilon\ll1$,
which is the only physically relevant regime from the point of view
of the slow envelope description of the weakly nonlinear light propagation
in FBG. }

The remainder of this paper is organized as follows. The family of
uniform solutions of the NLCME is briefly described in the next section
and its linear stability characteristics against perturbations with
wavelength $\lambda\sim1$ (transport scale) and $\lambda\sim\sqrt{|\varepsilon|}\ll1$
(dispersive scale) are obtained in sections 3 and 4, respectively.
These results indicate that the dispersive scales can destabilize
the uniform solutions for parameter values where the NLCME predict
stability. In the final section, numerical simulations of the NLCMEd
with small dispersion are used to corroborate the theoretical stability
predictions and some concluding remarks are also drawn.

\section{Continuous waves}

The constant modulus solutions of the NLCME, also known as continuous
waves (CW), can be written as \cite{deSterke98,KivsharAgrawal03}\textcolor{black}{\begin{eqnarray*}
 & A_{\textrm{cw}}^{+}=\rho\cos\theta\:\textrm{e}^{\textrm{i}\alpha t+\textrm{i}mx},\\
 & A_{\textrm{cw}}^{-}=\rho\sin\theta\:\textrm{e}^{\textrm{i}\alpha t+\textrm{i}mx},\end{eqnarray*}
where} $\rho>0$ is the power flowing through the grating, $\theta\in]-\frac{\pi}{2},0[\cup]0,\frac{\pi}{2}[$
measures the ratio between the two counterpropagating wavetrains,
and $\alpha$ and $m$ are given by\textcolor{black}{\begin{eqnarray*}
 &  & \alpha=\dfrac{\kappa}{\sin2\theta}+\dfrac{1+\sigma}{2}\rho,\\
 &  & m=(\dfrac{\kappa}{\sin2\theta}+\dfrac{1-\sigma}{2}\rho^{2})\cos2\theta.\end{eqnarray*}
According to eq. (\ref{Efield}), the associated electric field inside
the grating consists of the superposition of} two uniform wavetrains,
and $\alpha$ and $m$ \textcolor{black}{represent small corrections
to the frequency and wavenumber of the wavetrains}. The resulting
pattern resembles a standing wave for $\theta$ near $\pm\frac{\pi}{4}$
and a traveling wave for $\theta$ close to $0$ and $\pm\frac{\pi}{2}$\textcolor{black}{. }

In the linear limit of small light intensity, $\rho\rightarrow0$,
the relation between the frequency, $\alpha$, and the wavenumber,
$m$, of the CW takes the form \[
m^{2}=\alpha^{2}(1-\frac{\kappa^{2}}{\alpha^{2}}),\]
\begin{figure}
\begin{center}\includegraphics[%
  scale=0.4]{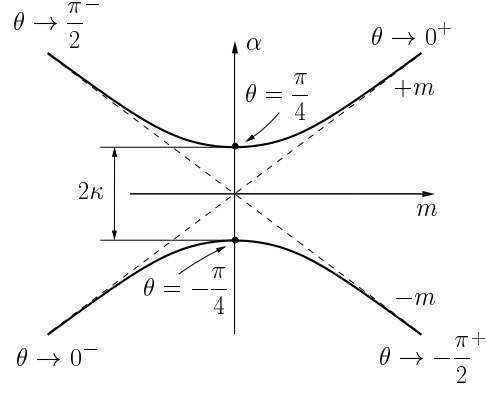}\end{center}

\caption{$\alpha-m$ plot of the CW in the linear regime $\rho\rightarrow0$.}
\end{figure}
which is plotted in Fig. 1 and shows a frequency gap of size $2\kappa$
where light propagation is not allowed (this is a typical resonance
effect in wave propagation in periodic media). As the power is increased,
the nonlinearity of the problem comes into play and the relation between
$\alpha$ and $m$ becomes more involved \[
m^{2}=(\alpha-\sigma\rho^{2})^{2}(1-\frac{\kappa^{2}}{(\alpha-(\frac{1+\sigma}{2})\rho^{2})^{2}}).\]
There are still two CW for any given $m$ if $\rho^{2}<\rho_{c}^{2}=2\kappa/(1-\sigma)$,
see Fig. 2a, while for $\rho^{2}>\rho_{c}^{2}$ a region of higher
multiplicity of solutions (up to 4) develops near $m=0$ \cite{KivsharAgrawal03},
as it can be appreciated from Fig. 2b . Notice that the frequency
gap persists for $\rho>0$ and it remains of size $2k$ but it is
shifted upwards by the nonlinear effects.%
\begin{figure}
\begin{center}\includegraphics[%
  scale=0.4]{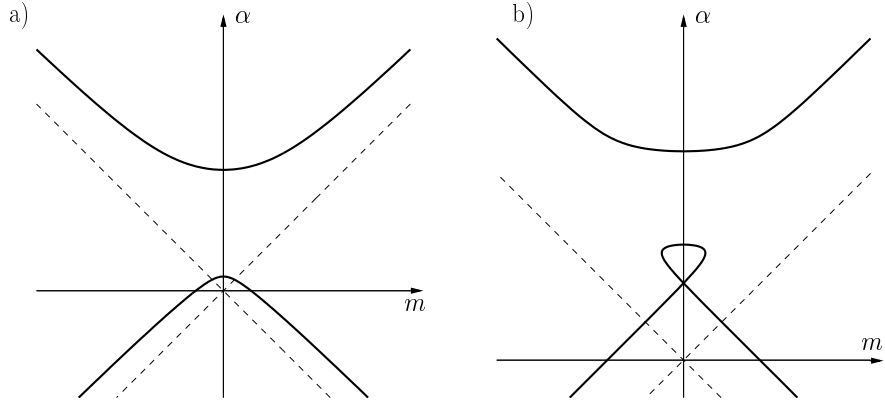}\end{center}

\caption{$\alpha-m$ plot of the CW for a) $\rho<\rho_{c}$ and b) $\rho>\rho_{c}$
($\sigma=\frac{1}{2}$).}
\end{figure}

In the rest of this paper the family of the CW is going to be represented
in the $(\theta,\rho^{2})$ plane shown in Fig. 3. The CW inside the
region marked by thick line \begin{equation}
\rho^{2}=\rho_{\times}^{2}=-\frac{2\kappa}{(1-\sigma)\sin(2\theta)}\label{ro2x}\end{equation}
 are those along the loop in Fig. 2b. The wavenumber $m$ is positive
(negative) in the white (shaded) areas and the thin lines correspond
to the values of the wavenumber compatible with the periodicity boundary
conditions (\ref{dispcc}), $m=2\pi n$ with $n\in\mathbb{Z}$ (arrows
indicate the $m$ increasing direction). Note that the $\theta=\pm\frac{\pi}{4}$
symmetry is due to the fact that the CW associated with $\theta>0$
($\theta<0$) and $\frac{\pi}{2}-\theta$ ($-\frac{\pi}{2}-\theta$)
are essentially the same after applying the spatial reflection symmetry
of the system $x\rightarrow-x\quad\textrm{and}\quad A^{+}\leftrightarrow A^{-}$,
\begin{figure}
\begin{center}\includegraphics[%
  scale=0.8]{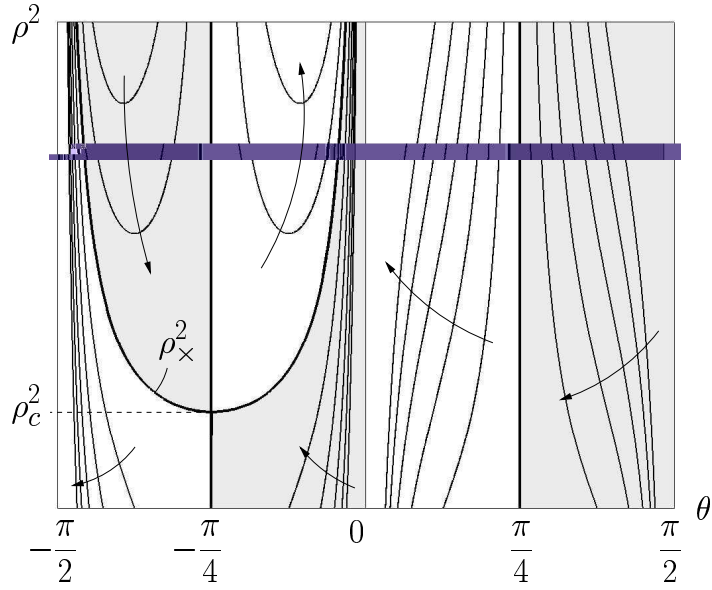}\end{center}

\caption{$(\theta,\rho^{2})$ representation of the CW ($\sigma=\frac{1}{2}$).}
\end{figure}

\section{CW stability: transport scale perturbations }

The CW are approximate solutions (up to order $\varepsilon$ corrections)
of the NLCMEd and its infinitesimal perturbations\[
A^{+}=A_{\textrm{cw}}^{+}(1+a^{+}),\qquad A^{-}=A_{\textrm{cw}}^{-}(1+a^{-}),\qquad\mbox{with}\quad|a^{\pm}|\ll1,\]
evolve according to the linearized version of the NLCMEd\begin{eqnarray}
a_{t}^{+}-a_{x}^{+} & = & \,\;\textrm{i}\kappa(a^{-}-a^{+})\tan\theta+\textrm{i}\sigma\rho^{2}\cos^{2}\theta(a^{+}+\overline{a^{+}})+\textrm{i}\rho^{2}\sin^{2}\theta(a^{-}+\overline{a^{-}})+\textrm{i}\varepsilon a_{xx}^{+},\label{pdepert+}\\
a_{t}^{-}+a_{x}^{-} & = & \textrm{i}\kappa(a^{+}-a^{-})/\tan\theta+\textrm{i}\sigma\rho^{2}\sin^{2}\theta(a^{-}+\overline{a^{-}})+\textrm{i}\rho^{2}\cos^{2}\theta(a^{+}+\overline{a^{+}})+\textrm{i}\varepsilon a_{xx}^{-},\label{pdepert-}\\
 &  & a^{\pm}(x+1,t)=a^{\pm}(x,t),\label{pdepertcc}\end{eqnarray}
 which, by means of the spatial Fourier expansion \[
(a^{+},a^{-})=\sum_{k=-\infty}^{\infty}(a_{k}^{+}(t),a_{k}^{-}(t))\:\textrm{e}^{\textrm{i}2\pi kx},\]
can be turned into a system of ordinary differential equations of
the form\begin{eqnarray}
\frac{da_{k}^{+}}{dt} & = & \textrm{i}(2\pi k)a_{k}^{+}+\textrm{i}\kappa(a_{k}^{-}-a_{k}^{+})\tan\theta+\textrm{i}\sigma\rho^{2}\cos^{2}\theta(a_{k}^{+}\!+\overline{a_{-k}^{+}})\nonumber \\
 &  & +\textrm{i}\rho\sin^{2}\theta(a_{k}^{-}+\overline{a_{-k}^{-}})-\textrm{i}\varepsilon(2\pi k)^{2}a_{k}^{+},\label{pert+}\\
\frac{da_{k}^{-}}{dt} & = & -\textrm{i}(2\pi k)a_{k}^{-}+\textrm{i}\kappa(a_{k}^{+}-a_{k}^{-})/\tan\theta+\textrm{i}\sigma\rho^{2}\sin^{2}\theta(a_{k}^{-}+\overline{a_{-k}^{-}})\nonumber \\
 &  & +\textrm{i}\rho^{2}\cos^{2}\theta(a_{k}^{+}+\overline{a_{-k}^{+}})-\textrm{i}\varepsilon(2\pi k)^{2}a_{k}^{-},\label{pert-}\end{eqnarray}
where $k\in\mathbb{Z}$ indicates the wavenumber of the perturbation.

Perturbations with wavelength of the order of the transport scale
have wavenumbers $k\sim1$ and for them the dispersion terms $\textrm{i}\varepsilon(2\pi k)^{2}a_{k}^{\pm}$
in the equations above are small and can be neglected in first approximation.
The resulting equations for $a_{k}^{\pm}$ together with those corresponding
to $\overline{a_{-k}^{\pm}}$ form a linear quartet uncoupled from
the rest whose solutions are of the form\begin{equation}
(a_{k}^{\pm}(t),\overline{a_{-k}^{\pm}(t)})=(a_{k}^{0\pm},\overline{a_{-k}^{0\pm}})\textrm{e}^{\textrm{i}\omega t},\label{eiomega}\end{equation}
 where $\omega$ is given by the following fourth order polynomial
with real coefficients\begin{eqnarray}
 & (\omega^{2}-(2\pi k)^{2})^{2}-2\kappa^{2}(\omega^{2}-(2\pi k)^{2})+4\kappa\rho^{2}\dfrac{\tan\theta}{1+\tan^{2}\theta}((2\pi k)^{2}(1+\sigma)-\omega^{2}(1-\sigma))\nonumber \\
 & -\kappa^{2}\tan^{2}\theta(\omega+(2\pi k))^{2}-\dfrac{\kappa²}{\tan^{2}\theta}(\omega-(2\pi k))^{2}=0.\label{poli-omega}\end{eqnarray}
This polynomial results from the NLCME, that is, when dispersion is
not considered, and it was previously analyzed in \cite{deSterke98}
for some particular values of $\rho^{2}$ and $\theta$. In this section
we will complete these results and give a detailed description of
the CW instability regions given by (\ref{poli-omega}). 

For the particular case of uniform perturbations, $k=0,$ the solutions
of (\ref{poli-omega}) can be calculated explicitly\begin{eqnarray*}
 & \omega=0\;\text{(double)}\quad\text{and}\\
 & \omega=\pm\sqrt{\dfrac{4\kappa^{2}}{\sin^{2}(2\theta)}+2\kappa\rho^{2}(1-\sigma)\sin(2\theta)},\end{eqnarray*}
and we have instability, i.e., $\omega$ with negative imaginary part,
when \begin{equation}
\rho^{2}>\rho_{0}^{2}=-\frac{2\kappa}{(1-\sigma)\sin^{3}(2\theta)}.\label{ro20}\end{equation}
This instability region lies inside the higher multiplicity region
given by eq. (\ref{ro2x}), see Fig. 4, and it destabilizes the CW
on the upper part of the loop in Fig. 2b. %
\begin{figure}
\begin{center}\includegraphics[%
  scale=0.6]{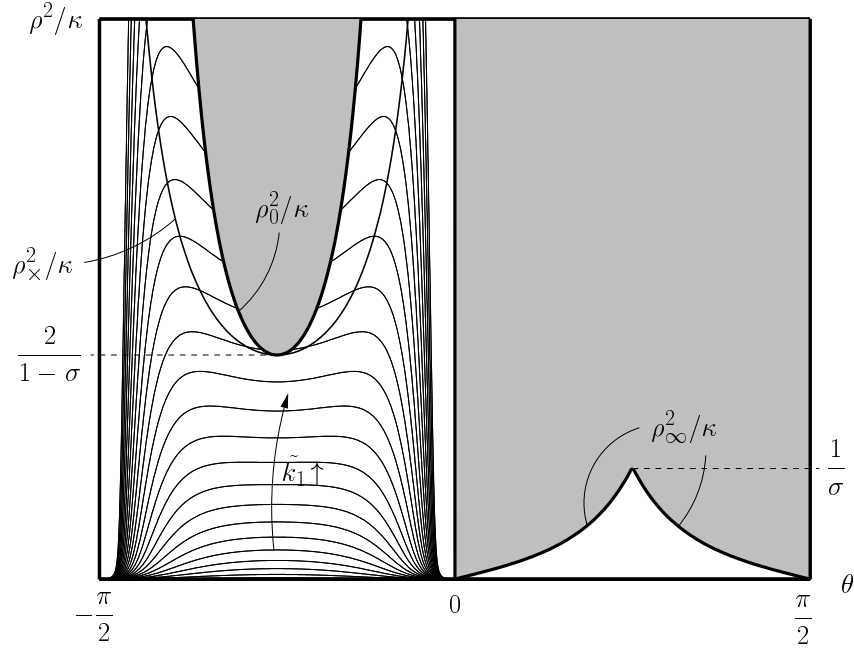}\end{center}

\caption{CW stability properties ($\sigma=\frac{1}{2}$). Shading indicates
instability and the thin lines correspond to $\tilde{k}_{1}=0.25+0.25n$,
with $n=0,1,2,\ldots$ . The CW above the $\tilde{k_{1}}=2\pi/\kappa$
line are unstable, see eq. (\ref{k1gorro}). }
\end{figure}

In the limit of large wavenumbers, $k\rightarrow\pm\infty$, the solutions
of eq. (\ref{poli-omega}) can be expanded as\[
\omega=\omega_{0}+\omega_{1}+\dots,\quad\text{with}\quad|\omega_{0}|\gg|\omega_{1}|\gg\dots.\]
The first order term $\omega_{0}\sim k\gg1$ satisfies the equation
\[
(\omega_{0}^{2}-(2\pi k)^{2})^{2}=0,\]
whose solutions are of the form $\omega_{0}\textrm{=}\pm2\pi k$ and
represent pure oscillations. The next order correction $\omega_{1}\sim1$
is given by\[
\omega_{1}^{2}=\kappa^{2}\tan^{\pm2}\theta-2\sigma\kappa\rho^{2}\dfrac{\tan\theta}{1+\tan^{2}\theta}\]
and the CW are therefore unstable inside the region defined by the
conditions\begin{equation}
\rho^{2}>\rho_{\infty}^{2}=\frac{\kappa\tan^{\pm2}\theta}{\sigma\sin(2\theta)},\label{ro2infty}\end{equation}
which is plotted in Fig. 4. Notice that, for the CW in this region,
all perturbations with wavenumber above a given one are unstable.
This unphysical behavior (amplification of perturbations far away
from the Bragg resonance) was already detected in \cite{deSterke98}
where it was suggested that it could be fixed by the addition dispersion.
This affirmation is true, as it will be seen in the next section where
the stability of perturbations with high wavenumbers is analyzed. 

The localization of the unstable perturbations with finite nonzero
wavenumber $0<|k|<\infty$ requires to numerically explore the roots
of (\ref{poli-omega}) and can be simplified by taking into account
the invariance of eq. (\ref{poli-omega}) under the transformations
\[
(\omega,k)\rightarrow(-\omega,-k)\quad\text{and}\quad(\omega,\tan\theta)\rightarrow(-\omega,1/\tan\theta)\]
that allows us to restrict the search to the parameter range\[
k>0\quad\text{and}\quad\theta\in[-\frac{\pi}{4},\frac{\pi}{4}].\]
Furthermore, if we use the scaled variables\begin{equation}
\tilde{\omega}=\omega/\kappa,\quad\tilde{k}=(2\pi k)/\kappa\quad\text{and}\quad\tilde{\rho}^{2}=\rho^{2}/\kappa,\label{kwrho}\end{equation}
 then the strength of the grating, $\kappa>0$, can be absorbed and
eq. (\ref{poli-omega}) can be rewritten as\begin{eqnarray*}
 & (\tilde{\omega}^{2}-\tilde{k}^{2})^{2}-2(\tilde{\omega}^{2}-\tilde{k}^{2})+4\tilde{\rho}^{2}\dfrac{\tan\theta}{1+\tan^{2}\theta}(\tilde{k}^{2}(1+\sigma)-\tilde{\omega}^{2}(1-\sigma))\\
 & -\tan^{2}\theta(\tilde{\omega}+\tilde{k})^{2}-\dfrac{1}{\tan^{2}\theta}(\tilde{\omega}-\tilde{k})^{2}=0.\end{eqnarray*}
The onset of complex roots (instability) takes place when both the
equation above and its derivative with respect to $\tilde{\omega}$
vanish. From these two conditions the values of $\theta$ and $\tilde{\rho}^{2}$
can be numerically computed for any given $\tilde{\omega}$ and $\tilde{k}$,
and the following results are obtained: 

\begin{itemize}
\item If $\theta>0$ then there are no more instabilities apart from the
one already obtained for $k\rightarrow\pm\infty$ (\ref{ro2infty}),
see Fig. 4.
\item For each negative $\theta$ and for each value of $\tilde{\rho}^{2}=\rho^{2}/\kappa$
there is a critical scaled wavenumber $\tilde{k}_{1}$ such that the
perturbations with $0<\tilde{k}<\tilde{k}_{1}$ are unstable and those
with $\tilde{k}>\tilde{k}_{1}$ are stable (constant $\tilde{k}_{1}$
curves are plotted as thin solid lines in Fig. 4). A CW will be then
unstable if the perturbation mode with lowest wavenumber ($k=1$)
is unstable, that is, if \begin{equation}
\frac{2\pi}{\kappa}<\tilde{k}_{1},\label{k1gorro}\end{equation}
recall the definition of $\tilde{k}$ in (\ref{kwrho}). For a given
value of $\kappa$, all the CW above the line $\tilde{k}_{1}=2\pi/\kappa$
in Fig. 4 are unstable.
\item The CW near the $\rho^{2}=\rho_{0}^{2}$ line exhibit a small interval
of stable scaled wavenumbers $\tilde{k}\in[0,\tilde{k}_{2}]$ with
$\tilde{k}_{2}<\tilde{k}_{1}$, but, as it has been checked numerically,
this does not modify the instability region defined above.
\end{itemize}
The compete stability characteristics of the CW against transport
scale perturbations (which are accurately described by the NLCME)
are summarized in Fig.4. There are three instability boundaries that
mark the destabilization of the uniform perturbations (\ref{ro20}),
high wavenumber perturbations (\ref{ro2infty}) and perturbations
with wavenumber $k=1$ (\ref{k1gorro}). The two first instability
conditions are not sensitive to the FBG length and depend only on
the combination $\rho^{2}/\kappa$ while the third one depends also
on $\kappa$ and it accounts for the stabilizing effect of the finite
size of the domain; in an infinite FBG ($\kappa\rightarrow\infty$,
see the last eq. in (\ref{escalados})) all CW with $\theta<0$ are
unstable due to (\ref{k1gorro}).

\section{CW stability: dispersive scale perturbations}

The CW perturbations with wavelength of the order of the dispersive
scale have large wavenumbers, $k\sim1/\sqrt{|\varepsilon|}\gg1$,
and for them the dispersion terms in eqs. (\ref{pert+})-(\ref{pert-})
can not be neglected. In order to study the stability of these perturbations,
we define the scaled wavenumber $K=(2\pi k)\sqrt{|\varepsilon|}\sim1$
and expand the solution of eqs. (\ref{pert+})-(\ref{pert-}) in the
form \[
a_{K}^{+}=a_{K0}^{+}(t,T)+\sqrt{|\varepsilon|}a_{K1}^{+}(t,T)+\cdots,\quad a_{K}^{-}=a_{K0}^{-}(t,T)+\sqrt{|\varepsilon|}a_{K1}^{-}(t,T)+\cdots,\]
where $T=\textrm{t}/\sqrt{|\varepsilon|}$ is a fast time scale. At
first order we obtain \begin{eqnarray*}
\frac{da_{K0}^{+}}{dT}-\textrm{i}Ka_{K0}^{+} & = & 0,\\
\frac{da_{K0}^{-}}{dT}+\textrm{i}Ka_{K0}^{-} & = & 0,\end{eqnarray*}
whose general solution is given by\[
(a_{K0}^{+},a_{K0}^{-})=(A_{K0}^{+}(t)\textrm{e}^{\textrm{i}KT},A_{K0}^{-}(t)\textrm{e}^{-\textrm{i}KT}).\]
Thus, in the fast time scale $T$, the group velocity term dominates
and these narrow dispersive perturbations simply travel in opposite
directions. At next order we obtain\begin{eqnarray*}
\frac{da_{K1}^{+}}{dT}-\textrm{i}Ka_{K1}^{+} & = & [-\frac{dA_{K0}^{+}}{dt}-\textrm{i}(\kappa\tan\theta+\dfrac{\varepsilon}{|\varepsilon|}K^{2})A_{K0}^{+}+\textrm{i}\sigma\rho^{2}\cos^{2}\theta(A_{K0}^{+}+\overline{A_{-K0}^{+}})]\textrm{e}^{\textrm{i}KT}\\
 &  & +[\textrm{i}\kappa\tan\theta A_{K0}^{-}+\textrm{i}\rho^{2}\sin^{2}\theta(A_{K0}^{-}+\overline{A_{-K0}^{-}})]\textrm{e}^{-\textrm{i}KT},\\
\frac{da_{K1}^{-}}{dT}+\textrm{i}Ka_{K1}^{-} & = & [-\frac{dA_{K0}^{-}}{dt}-\textrm{i}(\kappa/\tan\theta+\dfrac{\varepsilon}{|\varepsilon|}K^{2})A_{K0}^{-}+\textrm{i}\sigma\rho^{2}\sin^{2}\theta(A_{K0}^{-}+\overline{A_{-K0}^{-}})]\textrm{e}^{-\textrm{i}KT}\\
 &  & +[\textrm{i}\kappa/\tan\theta A_{K0}^{+}+\textrm{i}\rho^{2}\cos^{2}\theta(A_{K0}^{+}+\overline{A_{-K0}^{+}})]\textrm{e}^{\textrm{i}KT},\end{eqnarray*}
and the higher order correction $(a_{K1}^{+},a_{K1}^{-})$ will then
remain bounded in the fast time scale $T$ if the following equations
are satisfied\begin{eqnarray}
 &  & \frac{dA_{K0}^{+}}{dt}=-\textrm{i}(\kappa\tan\theta+\dfrac{\varepsilon}{|\varepsilon|}K^{2})A_{K0}^{+}+\textrm{i}\sigma\rho^{2}\cos^{2}\theta(A_{K0}^{+}+\overline{A_{-K0}^{+}}),\label{Ak+}\\
 &  & \frac{dA_{K0}^{-}}{dt}=-\textrm{i}(\kappa/\tan\theta+\dfrac{\varepsilon}{|\varepsilon|}K^{2})A_{K0}^{-}+\textrm{i}\sigma\rho^{2}\sin^{2}\theta(A_{K0}^{-}+\overline{A_{-K0}^{-}}),\label{Ak-}\end{eqnarray}
which give the evolution of $A_{K0}^{\pm}$ in the slow time scale
$t$. The left ($^{+}$) and right ($^{-}$) propagating perturbations
are not coupled, and the equations above together with the corresponding
ones for $\overline{A_{-K0}^{+}}$ and $\overline{A_{-K0}^{-}}$ form
two linear systems with constant coefficients whose solutions are
proportional to $\textrm{e}^{\Omega^{\pm}t}$, with\begin{eqnarray}
 &  & \Omega^{+}=\pm\sqrt{(\kappa\tan\theta+\dfrac{\varepsilon}{|\varepsilon|}K^{2})(2\sigma\rho^{2}\cos^{2}\theta-(\kappa\tan\theta+\dfrac{\varepsilon}{|\varepsilon|}K^{2}))}\quad\textrm{and}\label{omeplus}\\
 &  & \Omega^{-}=\pm\sqrt{(\kappa/\tan\theta+\dfrac{\varepsilon}{|\varepsilon|}K^{2})(2\sigma\rho^{2}\sin^{2}\theta-(\kappa/\tan\theta+\dfrac{\varepsilon}{|\varepsilon|}K^{2}))}.\label{omemenos}\end{eqnarray}
The dispersive perturbations are then unstable if the following conditions
are satisfied\begin{eqnarray*}
 &  & 0\leq\kappa\tan\theta+\dfrac{\varepsilon}{|\varepsilon|}K^{2}\leq2\sigma\rho^{2}\cos^{2}\theta,\\
 &  & 0\leq\kappa/\tan\theta+\dfrac{\varepsilon}{|\varepsilon|}K^{2}\leq2\sigma\rho^{2}\sin^{2}\theta,\end{eqnarray*}
that can be expressed as\begin{align}
 & \rho^{2}\ge\rho_{+}^{2}=\frac{\tan\theta}{\sigma\sin(2\theta)}(\kappa\tan\theta+\dfrac{\varepsilon}{|\varepsilon|}K^{2})\quad\,\,\,\,\,\text{with}\quad\tan\theta\ge\dfrac{\varepsilon}{|\varepsilon|}\frac{K^{2}}{\kappa},\label{rho2+disp}\\
 & \rho^{2}\ge\rho_{-}^{2}=\frac{\tan^{-1}\theta}{\sigma\sin(2\theta)}(\kappa\tan^{-1}\theta+\dfrac{\varepsilon}{|\varepsilon|}K^{2})\quad\text{with}\quad\tan^{-1}\theta\ge\dfrac{\varepsilon}{|\varepsilon|}\frac{K^{2}}{\kappa}.\label{rho2-disp}\end{align}
Notice that if we set $K^{2}=0$ in the equations above then the instability
condition given by eq. (\ref{ro2infty}) is recovered, in other words,
the limit $k\rightarrow\pm\infty$ for the $k\sim1$ (transport scale)
regime matches with the limit $K\rightarrow0$ for the $k\sim1/\sqrt{|\varepsilon|}\gg1$
(dispersive scale) regime. 

The above instability conditions depend on the parameter $K^{2}>0$
and define two families of regions in the $(\theta,\rho^{2})$ plane,
which are represented in Fig.5 for $\varepsilon>0$ and several values
of $K^{2}>0$ (note that expression (\ref{rho2+disp}) is identical
to (\ref{rho2-disp}) if we change $\tan\theta$ by $1/\tan\theta$
and therefore the instability region defined by the $\rho_{-}^{2}$
condition is simply the symmetric of that defined by $\rho_{+}^{2}$
around the vertical axes $\theta=\pm\dfrac{\pi}{4}$). A variation
of the wavenumber from $k$ to $k+1$ results in a very small increment
of the scaled wavenumber $\Delta K\sim\sqrt{|\varepsilon|}\ll1$,
so, in first approximation, we have to allow $K^{2}$ to vary continuously
between $0$ and $+\infty$ in (\ref{rho2+disp})-(\ref{rho2-disp}).
The resulting final dispersive instability region is then given by
the union of all these $K^{2}$ dependent regions. All CW with $\theta$
negative are thus rendered unstable for $\varepsilon>0$ (see Fig.5)
and, in a completely similar way, if $\varepsilon<0$ then all the
CW with positive $\theta$ are unstable (see Fig.6).

\begin{figure*}[!h]
\begin{center}\includegraphics[%
  clip,
  scale=0.6]{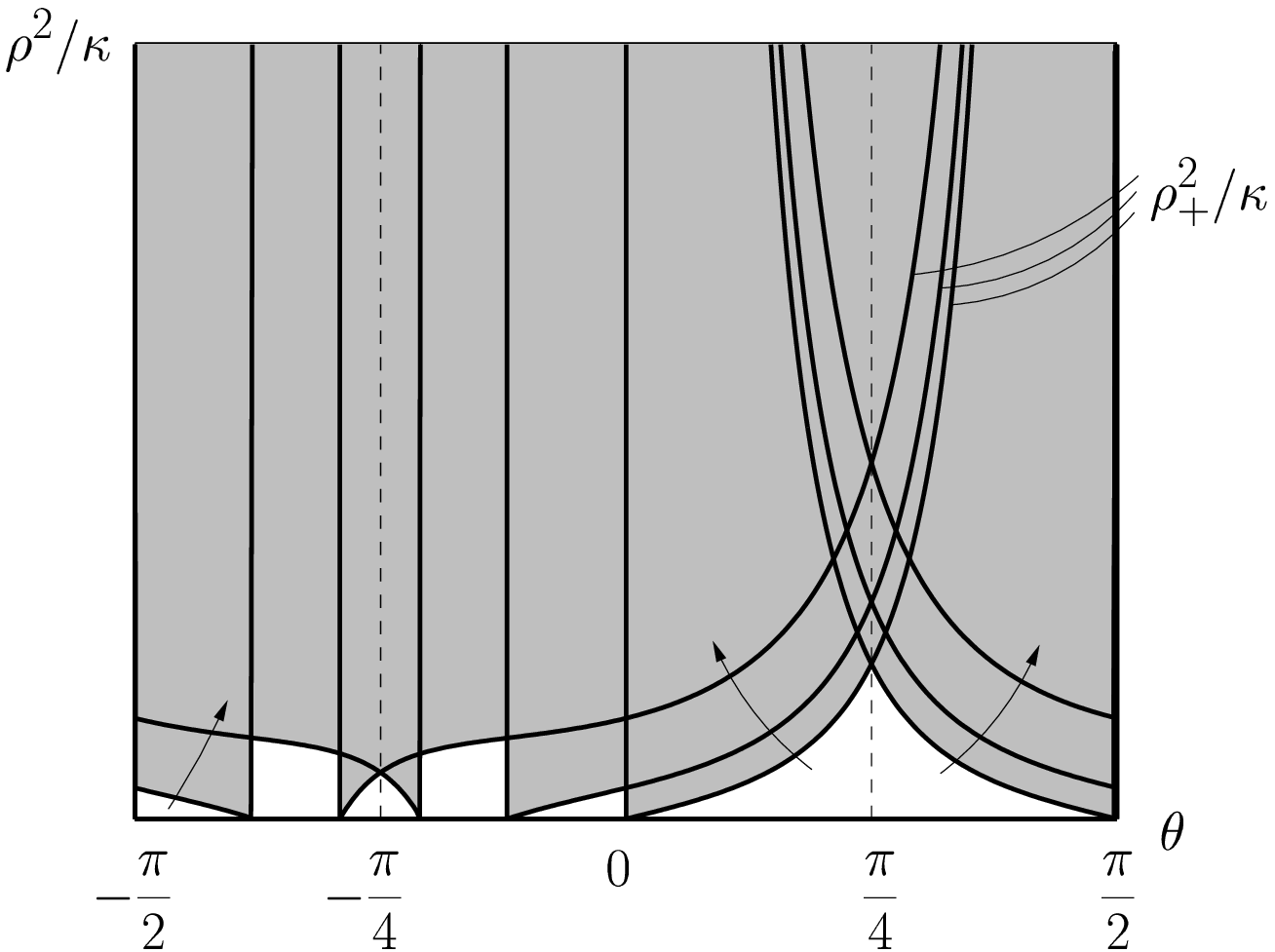}\end{center}

\caption{Dispersive instability regions defined by eqs. (\ref{rho2+disp})
and (\ref{rho2-disp}) for $\varepsilon>0$ and three different values
of $K^{2}$, with arrows pointing towards increasing $K^{2}$ direction
and shading indicating instability ($\sigma=\frac{1}{2}$).}
\end{figure*}

\begin{figure*}
\begin{center}\includegraphics[%
  clip,
  scale=0.6]{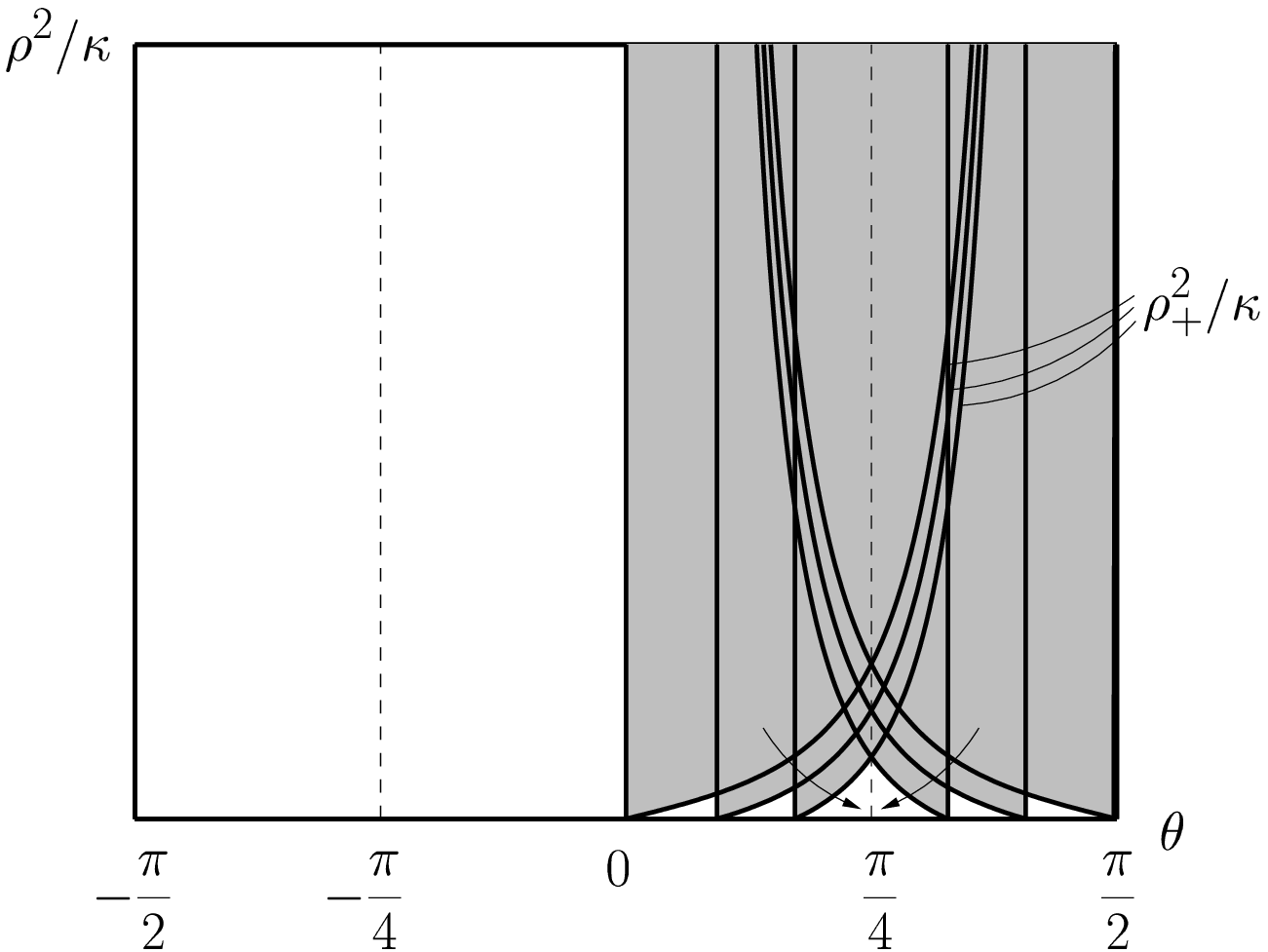}\end{center}

\caption{Dispersive instability regions defined by eqs. (\ref{rho2+disp})
and (\ref{rho2-disp}) for $\varepsilon<0$ and three different values
of $K^{2}$, with arrows pointing towards increasing $K^{2}$ direction
and shading indicating instability $(\sigma=\frac{1}{2}$).}
\end{figure*}

Therefore, the CW stability diagram represented in Fig.4 has to be
complemented with the following dispersive instability criterion:\begin{equation}
\-\textrm{-- for }\varepsilon>0\textrm{ (}\varepsilon<0\textrm{), all CW with }\theta<0\textrm{ (}\theta>0\textrm{) are unstable,}\label{dispinstcrit}\end{equation}
which states that all the CW along the lower (upper) branch in Fig.2
are unstable if $\varepsilon$ is positive (negative). 

As it was advanced in \cite{deSterke98}, the material dispersion
corrects the unphysical high wavenumber instability found in the previous
section, see eqs. (\ref{omeplus}) and (\ref{omemenos}) and the upper
plot in Fig.7. But it also gives rise to new instabilities (lower
plot of Fig.7) that are not detected if the NLCME description is used
and that are present independently of the dispersion sign: according
to (\ref{dispinstcrit}), positive (negative) dispersion is responsible
for the destabilization of NLCME stable CW with $\theta<0$ ($\theta>0$)
in the lower part of Fig.2. These high wavenumber dispersive perturbations
travel with the group velocity and their growth rate remain of order
unity as $\varepsilon\rightarrow0$, see eqs. (\ref{omeplus})-(\ref{omemenos}),
i.e., they evolve in the transport time scale $t\sim1$.

\begin{figure*}
\begin{center}\includegraphics[%
  clip,
  scale=0.5]{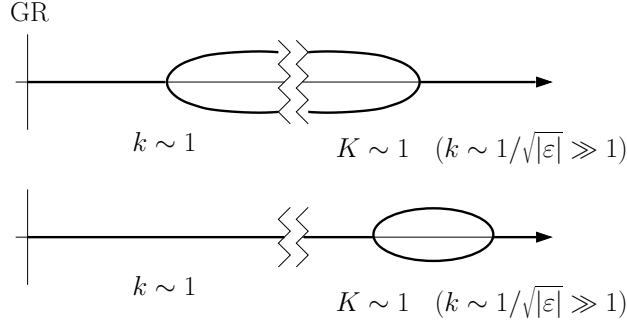}\end{center}

\caption{Sketch of the maximum growth rate of the transport and dispersive
perturbations as a function of the wavenumber}
\end{figure*}

\section{{\normalsize Discussion and conclusions}}

As a final check of the theoretical dispersive instability results
obtained in the previous section, several numerical simulations of
the NLCMEd have been performed for the two cases shown in Fig.8 with
different dispersion coefficients. The numerical method uses a Fourier
series in space with $N_{\textrm{Fourier}}$ modes and a 4th order
Runge-Kutta scheme for the time integration of the resulting system
of ODEs, with the linear diagonal terms integrated implicitly and
the nonlinear terms computed in physical space using the 2/3 rule
to remove the aliasing terms \cite{CanutoHussaniQuarteroniZang} (typical
required resolutions were in the range $N_{\text{F}}=256,\ldots,1024$
and $\Delta t=.01,\ldots,.001$). 

\begin{figure*}[!h]
\begin{center}\includegraphics[%
  clip,
  scale=0.9]{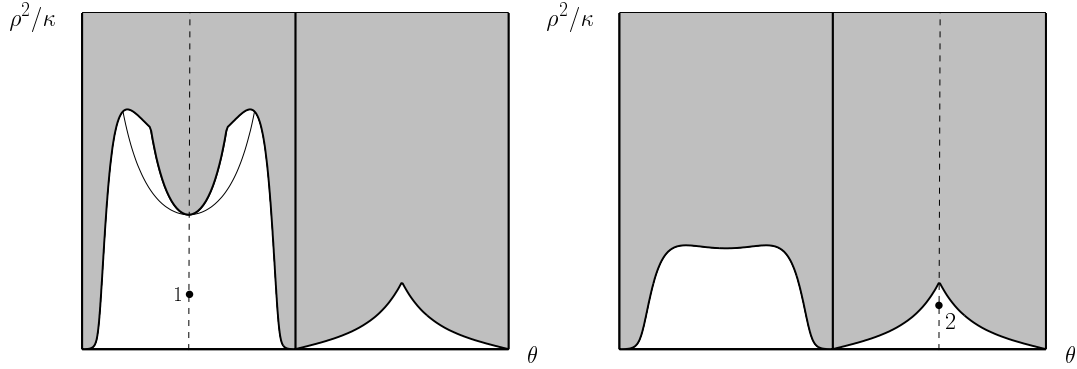}\end{center}

\caption{CW stability properties for $\kappa=1$ and $\kappa=2$, with shading
indicating instability (dispersive instabilities not shown) and dots
corresponding to CW1: $\kappa=1$, $\rho^{2}=1$, $\theta=-\frac{\pi}{4}$,
and CW2: $\kappa=2$, $\rho^{2}=2$, $\theta=\frac{\pi}{4}$ ($\sigma=\frac{1}{2}$).}
\end{figure*}

The CW1 is stable in the NLCME context (i.e., stable against transport
scale perturbations) and according to the dispersive instability criterion
(\ref{dispinstcrit}) is also stable for negative values of the dispersion.
This is corroborated by the numerical results presented in Fig.9,
where the time evolution of the spatial norm of the amplitudes and
its derivatives,\[
\| f\|=\sqrt{\int_{0}^{1}|f|^{2}\, dx},\]
 is plotted for $\varepsilon=-10^{-3}$. The CW1 remains stable and
the small initial perturbation does not grow. Note that the perturbation
cannot decay to zero because of the absence of dissipation in the
system, and thus the value of the spatial derivatives remains finite
but small (comparable to the size of the initial perturbation, see
Fig.9). In contrast, if dispersion is positive, the CW1 is unstable
due to the exponential amplification of the small dispersive scales,
see Fig.10. This growth is not a slow time effect; it takes place
in the time scale $t\sim1$ and does not vanish as $\varepsilon\rightarrow0$,
as it can be seen in the lower plot in Fig.10, where the evolution
of the spatial derivatives for two smaller dispersion values has been
added for comparison. The solution that develops is shown in Fig.11
and it consists of two counter-propagating wavetrains with dispersive
wavelength ($\sim\sqrt{\varepsilon}\ll1$) moving at the group velocity;
notice how the number of peaks is approximately doubled when the dispersion
is divided by 4. %
\begin{figure*}[!h]
\begin{center}\includegraphics[%
  clip,
  scale=0.8]{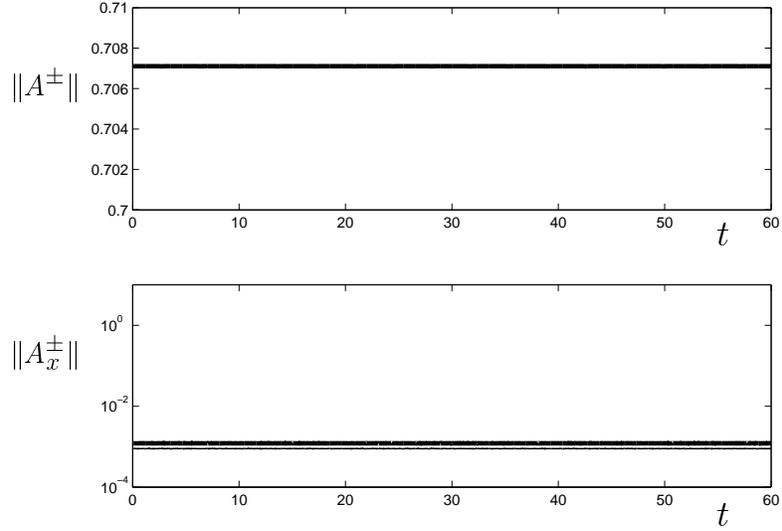}\end{center}

\caption{Thick (thin) lines indicate the time evolution of the spatial norm
of $A^{+}$ ($A^{-}$) and its spatial derivative for $\kappa=1$
and $\varepsilon=-10^{-3}$. The initial condition is the CW1 with
a random perturbation of size $\sim10^{-3}$ ($\sigma=\frac{1}{2}$). }
\end{figure*}

\begin{figure*}[!h]
\begin{center}\includegraphics[%
  clip,
  scale=0.8]{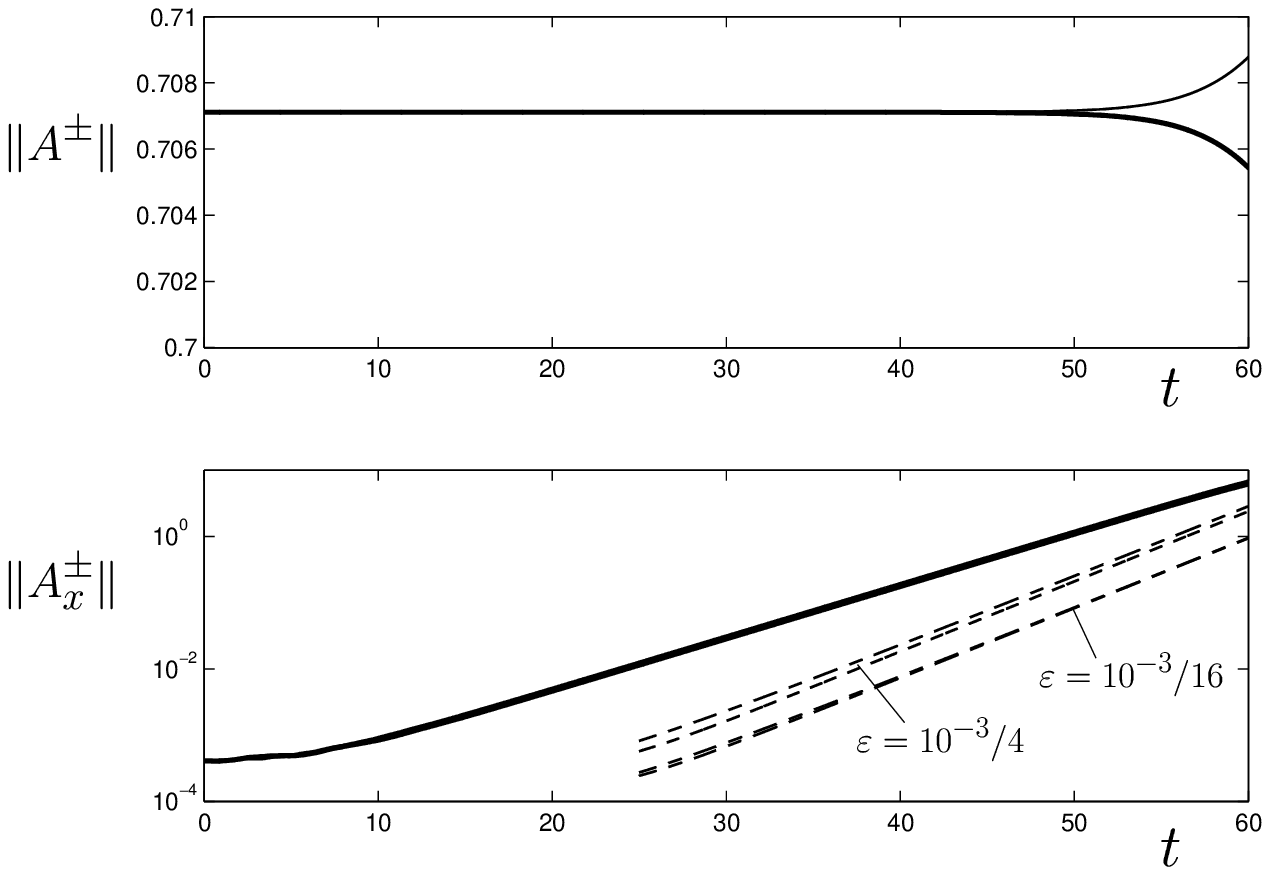}\end{center}

\caption{Thick (thin) lines indicate the time evolution of the spatial norm
of $A^{+}$ ($A^{-}$) and its spatial derivative for $\kappa=1$
and $\varepsilon=10^{-3}$ (dashed lines correspond to smaller dispersion
values). The initial condition is the CW1 with a random perturbation
of size $\sim10^{-3}$ ($\sigma=\frac{1}{2}$). }
\end{figure*}

\begin{figure*}[!h]
\begin{center}\includegraphics[%
  clip,
  scale=0.8]{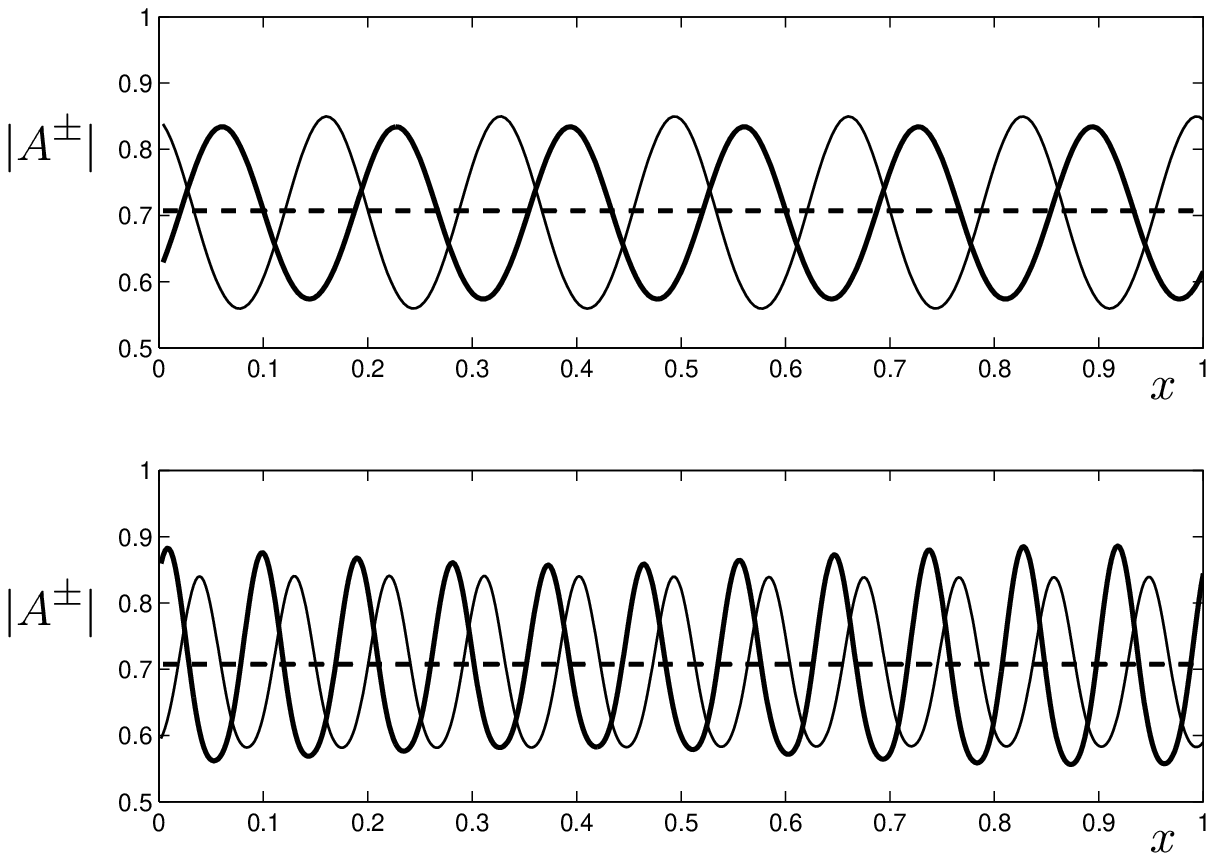}\end{center}

\caption{Snapshots of the solution of the NLCMEd at $t=60$ ($|A^{+}|$:thick
line, $|A^{-}|$:thin line), with parameters as in Fig. 10 and dispersion
$\varepsilon=10^{-3}$ (above) and $\varepsilon=10^{-3}/4$ (below). }
\end{figure*}

For the CW2 in Fig. 8 the situation is the opposite of that of the
CW1. Now the CW2 is stable for positive values of the dispersion (the
time evolution plot of the norms is omitted in this stable case because
it is identical to that shown in Fig.9) and the dispersive instability
develops, producing small dispersive scales all over the domain, for
negative dispersion, see Fig.12. The time plot of $||A_{x}^{\pm}||$
in Fig.12 shows the development of the dispersive instability for
several dispersion values; again, as predicted by the linear stability
theory, the dispersive instability exponent goes to a nonzero value
as the dispersion coefficient is reduced. A space-time representation
of the solution, once the dispersive scales are well developed, is
presented in Fig.13 where it can be clearly appreciated that, in the
short time scale $t\sim\sqrt{|\varepsilon|}$, the small dispersive
scales are just transported by the group velocity but they also evolve
on a slower time scale $t\sim1$ producing a very complicated spatio-temporal
pattern.

\begin{figure*}
\begin{center}\includegraphics[%
  clip,
  scale=0.8]{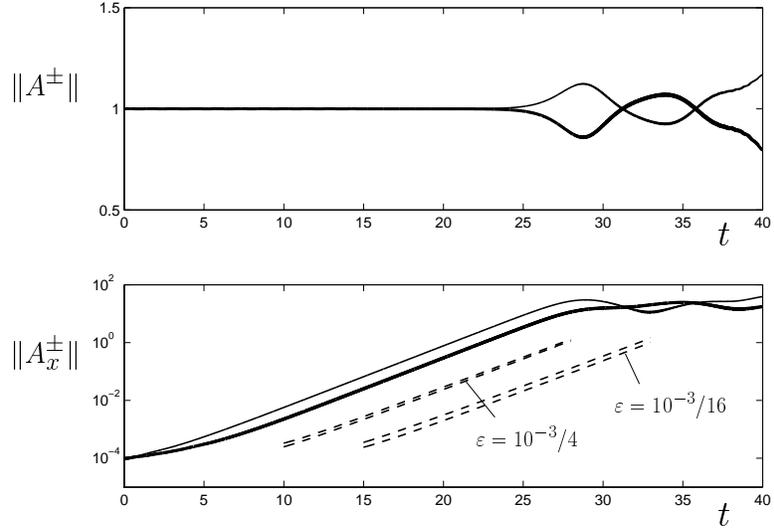}\end{center}

\caption{Thick (thin) lines indicate the time evolution of the spatial norm
of $A^{+}$ ($A^{-}$) and its spatial derivative for $\kappa=2$
and $\varepsilon=-10^{-3}$ (dashed lines correspond to smaller dispersion
values). The initial condition is the CW2 with a random perturbation
of size $\sim10^{-3}$ ($\sigma=\frac{1}{2}$). }
\end{figure*}
\begin{figure*}
\begin{center}\includegraphics[%
  clip,
  scale=0.8]{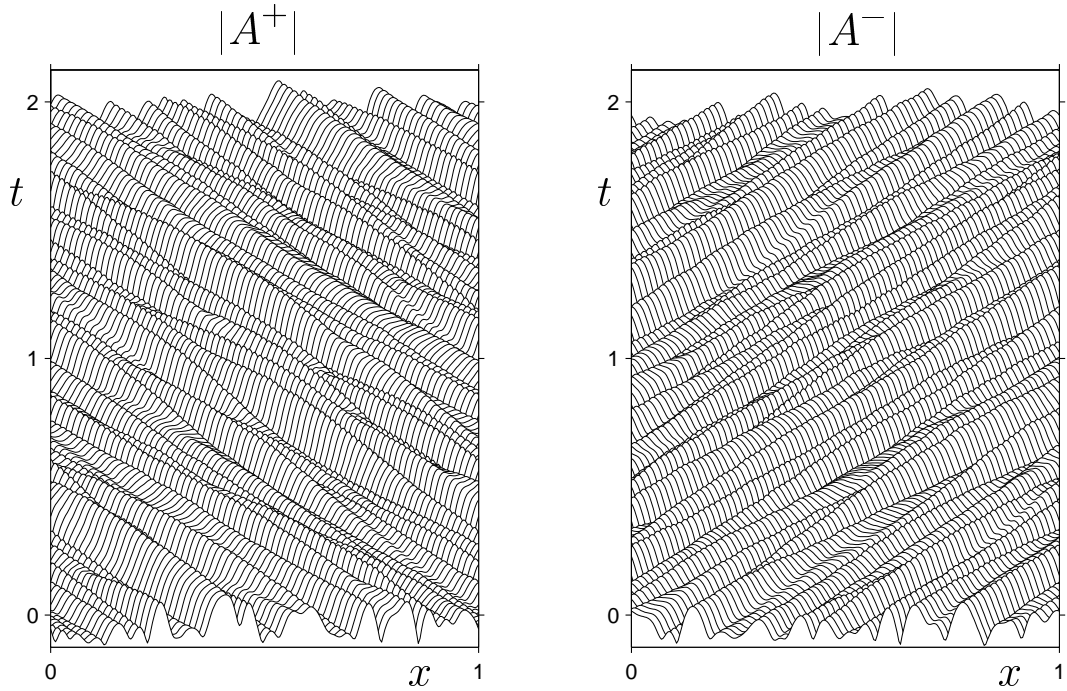}\end{center}

\caption{Space-time representation of the solution in Fig. 12 for two time
units after $t=80$. }
\end{figure*}

In conclusion, the main result of this paper is that the standard
NLCME formulation fails to predict the system dynamics and that, in
general, the small material dispersion terms should be taken into
account, i.e., the NLCMEd eqs. (\ref{NLCMEd+})-(\ref{NLCMEd-}) should
be used, to appropriately describe the weakly nonlinear regime of
light propagation in a FBG. In particular, the stability of the CW
is seen to be drastically affected by dispersion. No matter how small
is the dispersion or what sign it has, there are always stable CW
according to the NLCME formulation that are dispersively unstable.
The destabilization produced by the small dispersive terms is not
a higher order, longer time effect; it takes place in the time scale
of the NLCME and the associated growth rates remain of order one as
the dispersion coefficient goes to zero. Once the dispersive scales
are destabilized, they typically spread all over the domain producing
a very complicated spatio-temporal dynamics (see Fig. 13) that is
not captured by the NLCME. This behavior is the result of the competition
of two effects with different asymptotic order: the dominating advection
due to the group velocity and the dispersion. This situation generically
arises in any extended propagative system with spatial reflection
symmetry unless some special care is taken to reduce the group velocity.
It has been previously found in the description of double Hopf bifurcation
in dissipative systems \cite{martelvega96,martelvega98} and in parametrically
forced surface waves \cite{martelvegaknobloch03}.

There still remains the question of whether one should retain in the
envelope equations more terms of the form $\pm\alpha\varepsilon^{2}A_{xxx}^{\pm},\,\textrm{i}\beta\varepsilon^{3}A_{xxxx}^{\pm},\ldots$
and check if they produce new instabilities at even higher wavenumbers
$k\sim|\varepsilon|^{-2/3},|\varepsilon|^{-3/4},\ldots\gg|\varepsilon|^{-1/2}$.
This is not necessary as it can be readily seen as follows. The two
first dominant effects in the evolution of these very short wavelength
perturbations are transport and dispersion and, proceeding as in section
4, they can be written as\[
A_{k}^{\pm}(t)\textrm{e}^{\pm\textrm{i}kt+\textrm{i}\varepsilon k^{2}t+\textrm{i}kx}.\]
But, because of the effect of the dispersion, there is no possible
coupling between $A_{k}^{\pm}$ with $\overline{A_{-k}^{\pm}}$, which
is no longer resonant, and hence there is no possible exponential
instability (see eqs. (\ref{Ak+}) and (\ref{Ak-})) at these higher
wavenumbers.

As a final remark, it is interesting to mention that there are several
recent rigorous proofs \cite{GoodmanWeinsteinHolmes01,schneideruecker01}
that establish that the solutions of the NLCME remain asymptotically
close to solutions of the original physical equations. These proximity
results are not in contradiction with the results of this paper because
these proofs do not say anything about the stability of the solutions.
Therefore, when a stable NLCME solution is dispersively unstable,
its corresponding close solution of the physical equations is also
unstable and the system likes to move away from this solution and
develops dispersive scales, as the NLCMEd correctly predict. 

\begin{acknowledgments}
The author would like to thank Carlos M. Casas for his help with the
numerical simulations presented in this paper. This work was supported
by the European Office of Aerospace Research and Development under
contract FA8655-02-M4087 and by the Spanish Dirección General de Investigación
under grant MTM2004-03808.\bibliographystyle{apsrev}
\bibliography{/usuarios/primer/martel/TeX/chaos04/CHAOS04}

\end{acknowledgments}

\end{document}